\documentclass[10pt,aps,prd,preprintnumbers]{revtex4}
\usepackage{amsmath,amssymb}
\usepackage{graphicx}
\usepackage{subfigure}
\def\be{\begin{equation}}
\def\eq{\end{equation}}
\def\ee{\end{equation}}
\def\beq{\begin{equation}}
\def\eeq{\end{equation}}
\def\bea{\begin{eqnarray}}
\def\eea{\end{eqnarray}}
\def\no{\nonumber}
\def\wh{\widehat}

\begin{document}
\preprint{SU-4252-894}
\preprint{IMSc/2009/06/09}

\title{Beyond fuzzy spheres}

\author{T. R. Govindarajan $^{1,2}$\footnote{trg@imsc.res.in}, Pramod Padmanabhan $^{1,3}$
\footnote{ppadmana@syr.edu}, and T. Shreecharan $^1$\footnote{shreet@imsc.res.in}}
\affiliation{$^1$ The Institute of Mathematical Sciences, CIT Campus, Taramani, Chennai, India 600 113 \\
$^2$ Max Planck Institute for Gravitational Physics, D-14476 Golm, Germany\\
$^3$ Department of Physics, Syracuse University, Syracuse NY, 13244-1130, USA}

\begin{abstract}
We study polynomial deformations of the fuzzy sphere, specifically
given by the cubic or the Higgs algebra. We derive the Higgs algebra
by quantizing the Poisson structure on a surface in $\mathbb{R}^3$.
We find that several surfaces, differing by constants, are described
by the Higgs algebra at the fuzzy level. Some of these surfaces have
a singularity and we overcome this by quantizing this manifold using
coherent states for this nonlinear algebra. This is seen in the
measure constructed from these coherent states. We also find the
star product for this non-commutative algebra as a first step in
constructing field theories on such fuzzy spaces.
\end{abstract}

\maketitle

\section{Introduction}

Field theories on fuzzy spheres are being actively pursued in the
past few years \cite{Madore1,Bal1,Bal2,Bal3,Steinacker1,Bal-book}.
The primary interest in studying fuzzy spaces stems from the
attractive discretization it offers to regularize quantum field
theories preserving symmetries. It is known that a lattice
discretization though helpful in regularizing the field theory
breaks the symmetries of the theory under consideration and the full
symmetry is realized only when the regulator is taken to zero. This
problem does not occur in the fuzzy case. In the process the fermion
doubling problem is also avoided \cite{trg1,trg2,Immirzi}. Also the
possibility of new phases in the continuum theory which breaks
translation as well as other global symmetries can be studied
through simulations \cite{Ydri,Panero,Digal1,Digal2,Martin}.
Furthermore one can incorporate supersymmetry in a precise manner.
All these issues and many more attractive features can be found
discussed in detail in \cite{Bal-book}.

Fuzzy spheres and their generalizations were also considered in a
Kaluza Klein framework as extra dimensional space. There is also a
proposal for  dynamically generating such spaces in the same
framework \cite{Steinacker1,Madore2,Blando1}. Also analytically and
through simulations one can study evolution of geometries and their
transitions. Such spaces also arise as backgrounds in string theory
with appropriate Chern-Simons coupling \cite{Myers,Subrata}. Our
considerations here will be applicable in those scenarios too.

In the present work we go beyond the usual Lie algebra
characterizing the fuzzy sphere \be [X_j,X_k] = \frac{i\alpha
\epsilon_{jkl}}{\sqrt{N(N+2)}} X^l,~~\sum_i X_i^2~=~\alpha^2. \ee
What we have in mind is the study of various aspects of a surface
whose coordinates satisfy the cubic algebra, also known as  the
Higgs algebra (HA) \be [X_+, X_-] =  C_1 \, Z + C_2 \, Z^3,
~~[X_\pm,~Z]~=~\pm X_\pm. \ee This algebra originally arose  as a
symmetry algebra in the study of  the ``Kepler problem" in curved
spaces, particularly on a sphere \cite{Higgs,Leemon}. Quantum
mechanical Hamiltonian of a particle in a Kepler potential on the
surface of a sphere has a dynamical symmetry given by the above
algebra and can be used to solve the problem exactly.

The HA can be studied as a deformation of $SU(2)$ or
$SU(1,1)$. It can also be considered as a deformation of $SU_q(2)$ \cite{Zhedanov}. In this
sense the HA sits between the Lie and $q$ deformed algebras. Such
algebras are interesting not only from the physical point of view
but are also widely studied in mathematics \cite{Smith}.

In this paper we derive the Higgs algebra by quantizing the Poisson
structure on a manifold, which is embedded in $\mathbb{R}^3$. We
call this manifold, the Higgs manifold, $\mathcal{M}_H$, as the
algebra got from the Poisson algebra is the Higgs algebra. Arlind
et. al. \cite{Arlind} also produced new nonlinear deformations of
the $SU(2)$ algebra of different kinds which gave them torus
geometry as well as topology change. Another algebra that exhibits topology change is the Sklyanin algebra \cite{Skl}.

This paper is organized as follows; Sec. II shows the derivation of
the Higgs algebra from the Poisson algebra on the Higgs manifold.
The question of topology change in surfaces that are not round
spheres is also presented. In this process of topology change, we
encounter the singularity which is discussed in detail. In Sec. III,
we briefly introduce the interesting aspects of the HA and focus on
its finite dimensional representations. Sec. IV gives the
construction of the coherent states (CS) of the HA in detail.
Herein, we also provide the measure required for the resolution of
unity. These CS are then used in Sec. V to obtain the star product.
Our conclusions and outlook are presented in Sec. VI.

\section{The Higgs Manifold}

 We consider the following embedding in $\mathbb{R}^3$,
\be \label{surface}
x^2 + y^2 + (z^2 ~-~ \mu)^2 = 1.
\ee
This is the surface we call
as the Higgs manifold, $\mathcal{M}_H$. $\mu$ is a parameter which
can be varied. We now analyze this equation for different values of
$\mu$.

For $\mu = 1$, it is easy to see that there is a singular point
$(x=y=z=0)$ where the surface degenerates. But in the discrete case,
the representations do not display any difficulty at this value.
When $\mu < -1$ there are no solutions. For $-1 < \mu < 1$ we have a
deformed sphere, but still symmetric under rotations about the
z-axis. The surface becomes two disconnected spheres for $\mu > 1$.
These are explicitly shown in the figures below for specific values
of $\mu$.

\begin{figure*}
\begin{center}
\mbox{
    \leavevmode
    \subfigure [ \quad $\mu = 0$]
    { \label{f:subfig-1}
      \includegraphics[height=1.5in,width=1.5in]{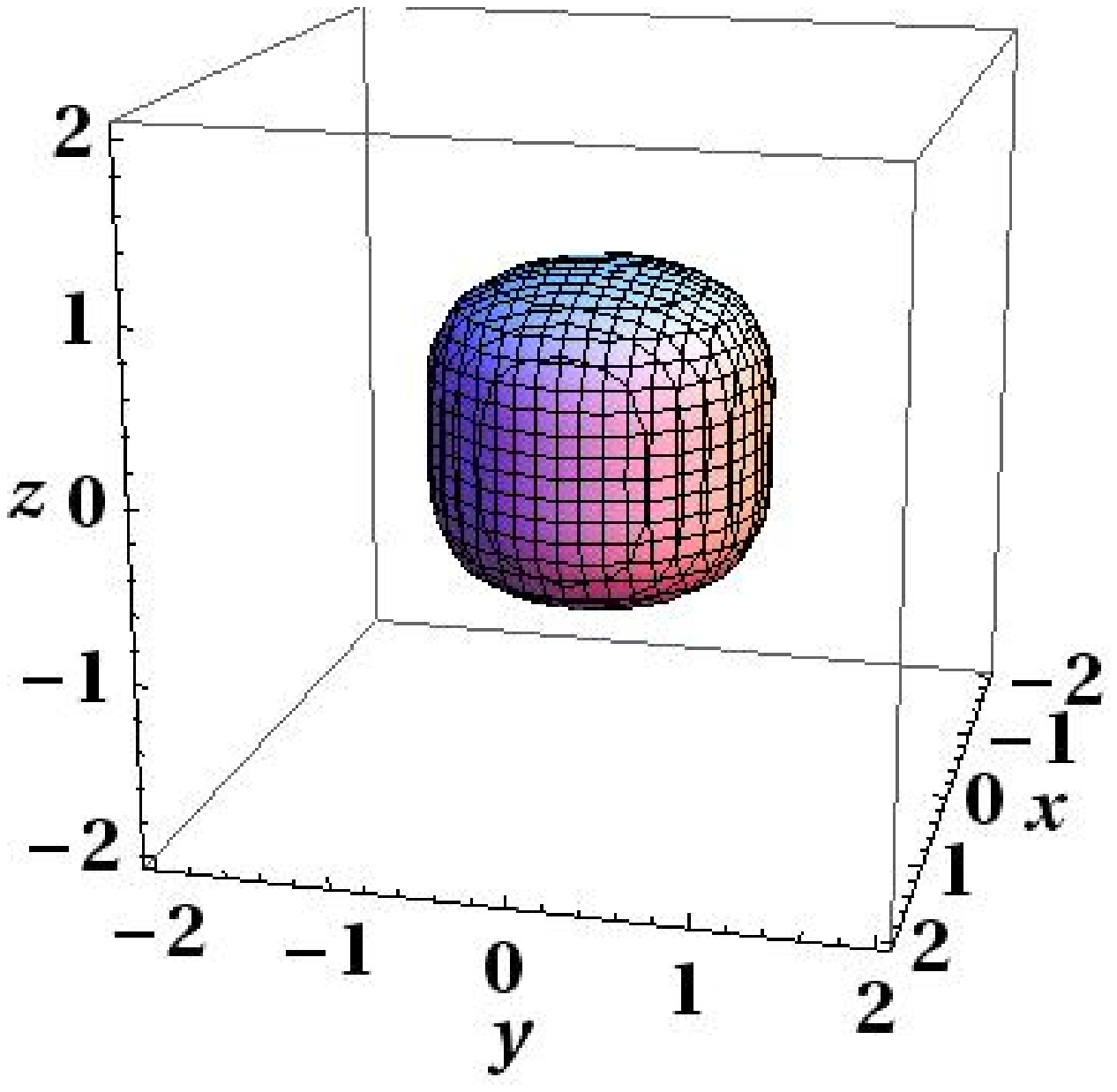} }

    \leavevmode
    \subfigure [ \quad $\mu = 1$]
    { \label{f:subfig-2}
      \includegraphics[height=1.5in,width=1.5in]{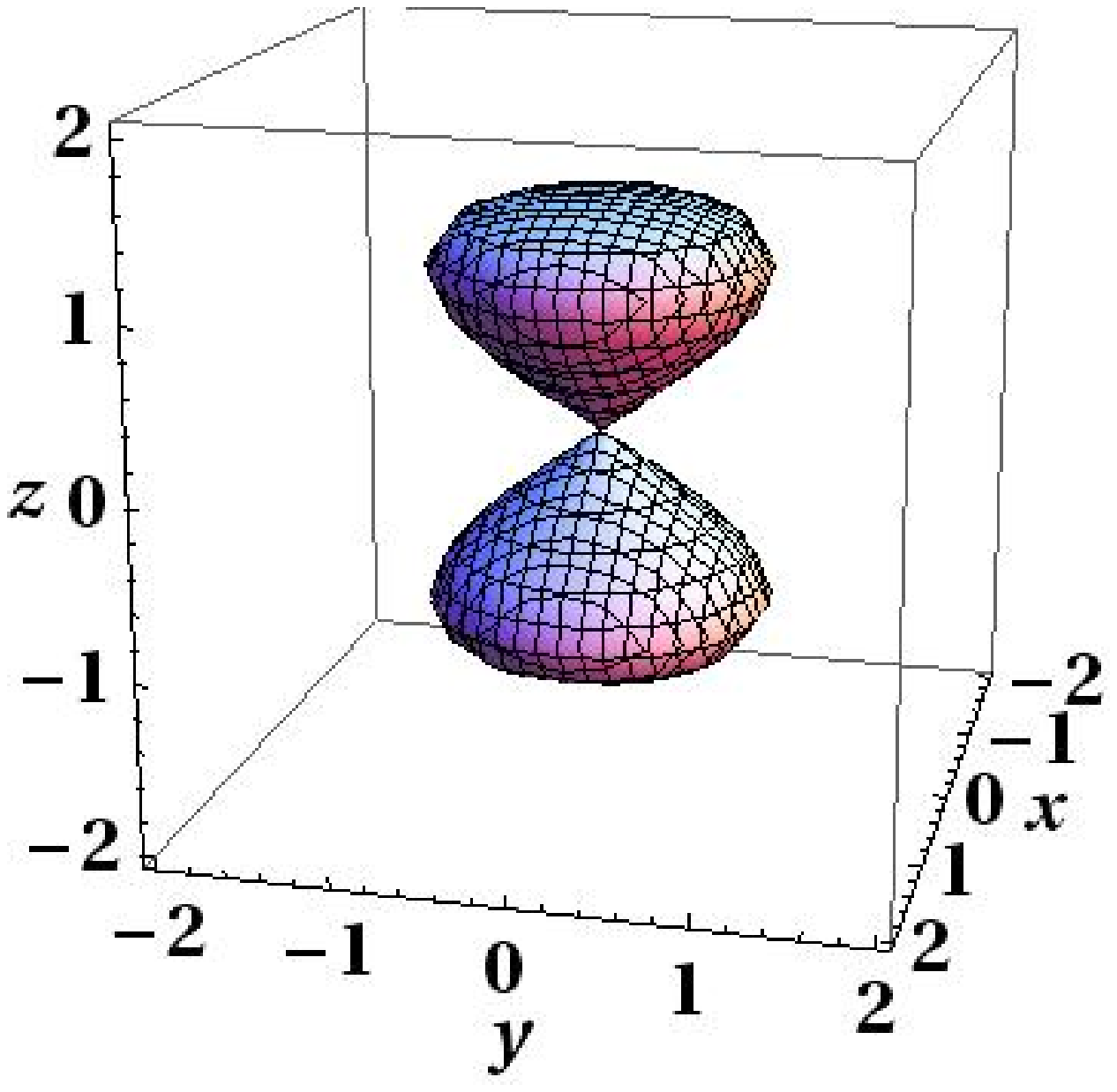} }
    \leavevmode
    \subfigure [\quad $\mu = 1.5$]
    { \label{f:subfig-3}
      \includegraphics[height=1.5in,width=1.5in]{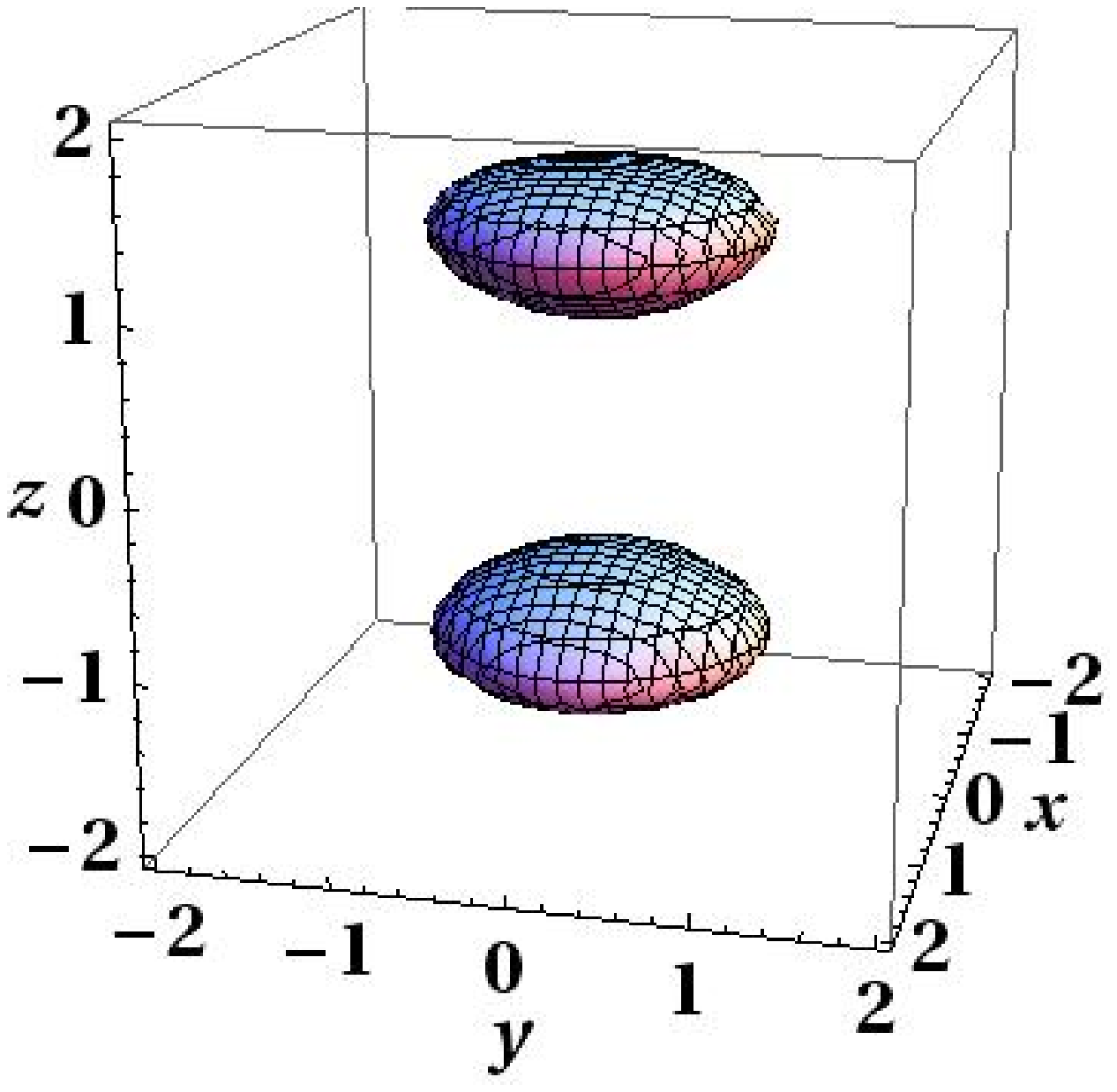} }}
\end{center}
\caption{Surface plots depicting the change in topology.}
\end{figure*}

\begin{center}
\textbf{The singularity}
\end{center}

 We use cylindrical coordinates to show the conical singularity
arising at $\mu=1$ and $x=y=z=0$ as shown in Fig. (1b). The equation
of the surface becomes \beq{r^2+(z^2-1)^2=1}.\eeq This acts as a
constraint giving $z$ in terms of $r$. Substituting this in the line
element of Euclidean $3$-space, in cylindrical coordinates,
$ds^2=dr^2+r^2d\phi^2+dz^2$, we get the induced metric on this
surface. \beq
ds^2=\left[1+\frac{r^2}{4(1-r^2)[1-\sqrt{1-r^2}]}\right]dr^2+r^2d\phi^2.
\eeq As $r\rightarrow 0$, the first term of the metric approaches
$\frac{3}{2}$. This implies a scaling of $r$ by
$\sqrt{\frac{3}{2}}$. This in turn induces a scaling of $\phi$ by
$\sqrt{\frac{2}{3}}$. Thus the new $\phi$ coordinate has range from
$0$ to $2\pi(1-\sqrt{\frac{2}{3}})$, making the origin a conical
singularity. This singularity cannot be removed by a coordinate
transformation.

\begin{center}
\textbf{The Poisson Algebra on $\mathcal{M}_H$}
\end{center}

We use the Poisson structure on $\mathbb{R}^3$ as defined by
\cite{Arlind}, to derive the Higgs algebra. This Poisson bracket is
given by
\be
\{f~g\} = \frac{\partial(C,f,g)}{\partial(x,y,z)}
\eq
where, $C=x^2 + y^2 + (z^2 ~-~ \mu)^2$ and $f$ and $g$ are two
functions on $\mathbb{R}^3$.

Using this we find the following: \be \{x~y\}=4z^3-4\mu z ,\ee \be
\{z~x\}=2y, \ee \be \{y~z\}=2x. \ee This, when quantized gives the
Higgs algebra.

\section{Higgs algebra and its representation}

We have shown in the previous section as to how the cubic Poisson bracket can be induced by a surface that is quartic in $z$. When we quantize this non-linear bracket we get the HA. The interest in studying these nonlinear algebras, apart from the physical applications \cite{Higgs,Leemon}, is that we can construct unitary finite or infinite dimensional representations. These and other interesting aspects were studied for many of these nonlinear structures, collectively called the polynomial algebras, by various authors \cite{Curtright,Rocek,Quesne,Floreanini,Beckers,Sunil}. In what follows we will explicitly state the representations of importance to us.

Let $X_+, X_-, Z$ be the generators of a three dimensional
polynomial algebra. This algebra is defined by the following
commutation relations:
\be
[X_+, X_-] =  C_1  Z + C_2  Z^3
\equiv f(Z),  \qquad [Z, X_\pm] = \pm  X_\pm.
\ee
In the above $C_1$
and $C_2$ are arbitrary constants. It is straight forward to check
that the Jacobi identity is preserved.  When $C_2 = 0$ and $C_1= 2$
or $C_1 = -2$, we have the $\mathfrak{su}(2)$ or
$\mathfrak{su}(1,1)$ algebra respectively. We will be interested in the cubic algbera that is treated as a deformation of the $\mathfrak{su}(2)$ algebra. Hence we will consider finite dimensional representations only.


The finite dimensional irreducible representations of the HA are characterized, like in $SU(2)$,
by an integer or half integer $j$ of dimension $2j~+~1$.
\bea
\no Z \, \vert j,m \rangle & = & m \, \vert j, m  \rangle  \ , \\
X_+ \, \vert j,m \rangle & = & \sqrt{g(j) - g(m)} \, \vert j, m + 1
\rangle \ . \eea The structure function $g(Z)$ is chosen such that
$f(Z) = g(Z) - g(Z-1)$. Note that $g(Z)$ is defined only upto the
addition of a constant. Later we will see this freedom plays an
importatnt role in arriving at the one parameter family of surfaces
namely Eq. (\ref{surface}). For arbitrary polynomials $f(Z)$, one
can solve and find solutions for $g(Z)$ \cite{Quesne}. The fact that
the we can write $f(Z)$ as difference of structure functions $g(Z)$
enables one to find the Casimir $\mathcal{C}$ of the algebra in an
almost trivial way. The Casimir $\mathcal{C}$ is: \be
\label{higgscas} \mathcal{C} = \frac{1}{2}\left[\{X_+, X_-\} + g(Z)
+ g(Z-1)\right], \ee \be \label{higgsrep} \mathcal{C} \, \vert j,m
\rangle  =  g(j) \, \vert j, m \rangle \ee where the curly brackets
denote anti-commutator. It is easy to verify
$[\mathcal{C},X_\pm]~=~[\mathcal{C},Z]~=~0$. So far we have not
specified what the explicit form of $g(Z)$ is and without further
ado we state for our case of HA: \be \label{commfunc} g(Z) = C_0 +
\frac{C_1}{2} \, Z (Z+1) + \frac{C_2}{4} \, Z^2 (Z+1)^2 . \ee Here
$C_0$ is a constant. Now the Casimir as a function of $Z$ alone
assumes a form of a single or double well potential depending on the
values of the parameters. The physical meaning of this behavior can
be understood from the work of Rocek \cite{Rocek}. The condition for
finite dimensional representations is also discussed in
\cite{Rocek}. In our case we note that $g(Z) = g(-Z-1)$, which is
also the condition for the case of the $SU(2)$ algebra. This makes
the function $g$ periodic and hence we can be sure that we have
finite dimensional representations for the choice of parameters we
will make for our Higgs algebra.

Applying Eq. (\ref{commfunc}) to Eq. (\ref{higgscas}) and then comparing it with Eq. (\ref{surface}), we get $C_0 = \mu^2$, $C_1=-2(2\mu~+~1)$, and $C_2=4$. Let us note that though there is a singularity in the continuum limit, in the
discrete case we have a valid representation theory as we vary the
parameters. This looks like a novel resolution of singularity. Similar behavior was noted in \cite{Arlind} where, the
topology changes from a sphere to a torus with a degenerate surface
at a transition point in the parameter space.

Now we will construct the CS for this nonlinear algebra to get a
better understanding of the semiclassical behavior.

\section{The Higgs algebra coherent states}

The field CS \cite{Glauber,Sudarshan} and their
generalizations \cite{Barut,Gilmore,Perelomov} been extensively
studied from various aspects, motivated mainly by applications to
quantum optics. But, we are interested in them as providing
appropriate semiclassical descriptions of the nonlinear algebra. As
is well known there are two types of CS. (1) those that
are ``annihilation operator'' eigenstates also known as
Barut-Girardello CS \cite{Barut} (2) states obtained
through the action of the displacement operator also known as
Perelemov states \cite{Perelomov}. The first is useful when
considering non compact groups like $SU(1,1)$ and the second for
compact ones.

We consider the finite dimensional representation of the Higgs algebra as we
want to view it as a deformation of the fuzzy sphere algebra. Hence,
we resort to the construction via the displacement operator. One
should keep in mind that since our algebra is nonlinear one cannot
attach any group theoretical interpretation to such states. The
actual procedure should be viewed as an algebraic construction and
has been carried out in \cite{Jagan,Sadiq}.

Since, the algebra under study is not a Lie algebra, a
straightforward application of the Perelomov prescription is also
not possible, wherein essential use of the Baker-Campbell-Hausdorff
(BCH) formula is made. To get around it we find a new operator
$\wh{X}_-$ such that $[X_+, \wh{X}_-] = 2 \, Z$. Let $\wh{X}_- = X_-
\, G(\mathcal{C},Z)$; substituting this in the commutator relation
we get
\be \label{su2su11}
X_+ \, X_- \, G(\mathcal{C},Z) - X_- \, X
_+ \, G(\mathcal{C},Z+1) = 2 \, Z \ .
\ee Choose the {\it ansatz} for
$G$ of the form
\be
G(\mathcal{C},Z) = \frac{- Z \, (Z+1) +
\lambda}{\mathcal{C} - g(Z-1)},
\ee
where $\lambda$ is an arbitrary
constant. Now that we have the `ladder' operators that obey the
$\mathfrak{su}(2)$ algebra, we can use the Perelomov prescription.
The CS are given by \be \vert \zeta \rangle = e^{\zeta X_+ \, - \,
\zeta^{\ast} \wh{X}_-} \vert j,-j \rangle. \ee Disentangling the
above exponential, using the BCH formula for $\mathfrak{su}(2)$ and
$\wh{X}_- \vert j,-j \rangle = 0$ we find the expression for the CS
acquires the form
\be \label{cshiggs}
\vert \zeta \rangle =
N^{-1}(|\zeta|^2) \, e^{\zeta X_+} \, \vert j,-j \rangle.
\ee
where $N^{-1}(|\zeta|^2)$ is the normalization constant that is yet to be
determined and $\zeta \in \mathbb{C}$. Notice that the ladder
operators that form the `Lie algebra' are not mutually adjoint. The
above state is to be viewed as ``non-linear $\mathfrak{su}(2)$
coherent state" and are very similar in spirit to the CS of
nonlinear oscillators \cite{Manko} and extensively used in quantum optics.

Now we will study whether the above definition of CS is
suitable. The requirements for $\vert \zeta \rangle$ to be CS have been enunciated
by Klauder \cite{Klauder}: (1) $\vert
\zeta \rangle$ should be normalizable, (2) $\vert \zeta \rangle$
should be continuous in $\zeta$, (3) $\vert \zeta \rangle$ should
satisfy resolution of identity. We will consider normalization and
resolution of identity in the following.

\subsection{Normalization}

To find the normalization constant $N^{-1}(|\zeta|^2)$ we compute the scalar product
of HACS and set it equal to $1$. We get
\bea \no
N^{2}(|\zeta|^2)
& = & \langle j,-j \vert e^{\bar{\zeta} X_-} \, e^{\zeta X_+} \vert j,-j\rangle \ , \\ \no
& = &  1+ \sum_{n=1}^{2j} \frac{|\zeta|^{2n}}{(n!)^2} \, \prod_{\ell=0}^{n-1}K_{j,-j + \ell} \, \prod_{p=0}^{n-1}H_{j,-j+n-p} \ , \\ \no
& = &  1 + \sum_{n=1}^{2j}\frac{|\zeta|^{2n}}{(n!)^2} \, \prod_{\ell=0}^{n-1}(K_{j,-j+\ell})^2 \ , \\ \no
& = &  1 + \sum_{n=1}^{2j}|\zeta|^{2n} \, \left(
{\begin{array}{*{10}c} 2j \\ n \\ \end{array}}\right) \\
&& \prod_{\ell=0}^{n-1}\left(\frac{C_1}{2}+\frac{C_2}{4}
[2j(j-\ell)-\ell(\ell+1)]\right).
\eea
In the above expression $K_{j,m} \equiv H_{j,m+1}=\sqrt{g(j)-g(m)}$. Observe that the
expression under the product is quadratic in $\ell$ and can be
factorized.
\bea \no
N^{2}(|\zeta|^2) & = & 1 + \sum_{n=1}^{2j}|\zeta|^{2n} \, \left( {\begin{array}{*{10}c} 2j \\ n \\ \end{array}}\right) \,
\prod_{\ell=0}^{n-1} (\ell-A_+)(\ell-A_-), \\
& = & 1 + \sum_{n=1}^{2j}|\zeta|^{2n} \, D_n,
\eea
where
\be
A_{\pm} = - \left[(j + \frac{1}{2}) \pm \sqrt{(j + \frac{1}{2})^2 +
(2 j^2 + \frac{2 C_1}{C_2} )} \right].
\ee
Taking the ratio of $D_{n+1}/D_n$ we get
\be \frac{D_{n+1}}{D_n}~=~\frac{(n-A_+)(n-A_-)(2j-n)}{(n+1)}.
\ee
It can be seen that this is the condition for the generalized hypergeometric series for
${_3}F_0(-A_+,-A_-,-2j;0;-|\zeta|^2)$.
Hence, the final expression for the normalization constant of the HA
is \be \label{normconst} N^{2}(|\zeta|^2) =
{_3}F_0(-A_+,-A_-,-2j;0;-|\zeta|^2) . \ee

\subsection{Resolution of identity}

The resolution of identity is one very important criterion that any CS must satisfy:
\be \label{resid}
\frac{1}{\pi}\int \vert \zeta \rangle d\mu(\zeta,\bar{\zeta}) \langle \zeta \vert = \mathbb{I}.
\ee
The integration is over the complex plane. Introducing the HACS in the above equation and writing the resulting equation in angular coordinates, $\zeta= r \, e^{i \theta}\, (0 \leq \theta < 2 \pi)$, brings us to
\be
\sum_{n=0}^{2j} \int dr \, \frac{\rho(r^2)}{N^2(r^2)} \frac{r^{2n+1}}{(n!)^2} \, X_+^n \vert j, -j \rangle \langle -j, j \vert \ X_-^n = \mathbb{I}.
\ee
We know that the angular momentum states are complete and hence for the above equality to hold the integral should be equal to one. Defining $\tilde{\rho}(r^2) \equiv \rho(r^2)/N^2(r^2)$ and simplifying the product as shown in the previous subsection we have
\be \label{residrad}
\int\limits_{0}^{\infty} dr \ r^{2n+1} \ \tilde{\rho}(r^2)  = \Gamma(n+1) \times
 \frac{\Gamma(A^\prime_+ - n + 1) \ \Gamma(A^\prime_- - n + 1) \Gamma(2j- n + 1)}{\Gamma(2j+1) \ \Gamma(A^\prime_+ + 1) \ \Gamma(A^\prime_- + 1)},
\ee
Where $A^\prime_{\pm} = -A_{\pm}$. Making a change of variable, $r^2 = x$ and replacing the discrete variable $n$ by the complex one $(s-1)$ we notice that the weight function $\tilde{\rho}(x)$ and the r.h.s. of the above equation become a Mellin
transform related pair \cite{Sixdeniers}. The unknown function $\tilde{\rho}(x)$ can be read of
from tables of Mellin transforms \cite{Oberhettinger}. For the sake of
completeness we reproduce the relevant formula below
\be \label{G290}
\int\limits_{0}^{\infty} dx \ x^{s-1} \ G_{p , \ q}^{m , \ n} \bigg(
x \bigg\vert \begin{array}{l}a_{1},\ldots, a_{p} \quad \\[2pt] b_{1},\ldots,
b_{q}, 0\end{array} \bigg)  =
\frac{\prod_{j=1}^{m} \Gamma(b_{j} + s)  \prod_{j=1}^{n} \Gamma(1-a_{j} - s)}{\prod_{j=m+1}^{q} \Gamma(1-b_{j} - s)  \prod_{j=n+1}^{p} \Gamma(a_{j} + s)}  ,
\ee
where the r.h.s.\ is the $s$-dependent part of of the weight
function. $G_{m,\ n}^{p, \ q}$ is called the
Meijer-G function and more details can be found in \cite{Mathai}. Casting
equation (\ref{residrad}) in the above standard form we find that
\be \label{weight}
 \rho(|\zeta|^2)  =  \frac{{_3}F_0(A^\prime_+,A^\prime_-,-2j;0;-|\zeta|^2)}{ \Gamma(2j+1) \Gamma(A_+^\prime + 1) \Gamma(A_-^\prime + 1)}
\ G^{1 , \ 3}_{0 , \ 0} \bigg[-|\zeta|^2 \bigg\vert \begin{array}{l}-2j-1, -(A^\prime_+ + 1), - (A^\prime_- + 1) \\[1pt] 0 \end{array} \bigg].
\ee
At $\mu=1$, this function is well behaved, leaving no trace of the
conical singularity encountered in the continuum. This can be seen
as a consequence of quantizing the Higgs manifold using the HACS.

\section{The star product}

The star product plays a ``stellar" role in the study of deformation
quantization and noncommutative geometry. There exist in the
literature a host of such products depending on various situations.
For example the Moyal product arises when one is dealing with a
noncommutative plane and the underlying algebra  happens to be the
Heisenberg-Weyl type. Similarly the Kontsevich product arises when
the noncommutative parameter itself, is a function of the
coordinates. We in this section are interested in constructing a
new product that will reduce to the star product of the fuzzy sphere
when the parameter pertaining to the cubic term is set to zero. The
technique for obtaining star product for the HA, follows
Grosse and Presnajder \cite{Grosse} and we refer to it for details
regarding the use of CS in this construction. Suffice it to mention
here that CS ensures that the product obtained is associative.

The algebra of functions on the Higgs manifold, $\mathcal{M}_H$, is commutative under point-wise
multiplication. When we quantize this manifold, this point-wise
product is deformed to an associative star product which is
noncommutative.

We consider the algebra of operators, $\mathcal{A}$, generators of which satisfy the HA.
These operators act on some Hilbert space. We then use the symbol
to map these operators to the functions on the Higgs manifold.
The symbol map is defined as follows
\beq
\phi :
\mathcal{A}\rightarrow\mathcal{M}_H \ .
\eeq
We use the HACS to define the symbol map in this case:
\bea
\phi(\hat{\alpha}) \equiv \langle\zeta|e^{\alpha_-X_-}e^{\alpha_0Z}e^{\alpha_+X_+}|\zeta\rangle
& = & N^{-2}\langle
j,-j|e^{\alpha_0Z}e^{-\alpha_0Z}e^{(\alpha_-+\zeta^*)X_-}e^{\alpha_0Z}e^{(\alpha_++\zeta)X_+}|j,-j\rangle \ ,
\\ \nonumber & = & N^{-2}e^{-j\alpha_0}\langle
j,-j|e^{(\alpha_-+\zeta^*)e^{\alpha_0}X_-}e^{(\alpha_++\zeta)X_+}|j,-j\rangle \ ,
\\ \nonumber & = & e^{-j\alpha_0}\frac{_3F_0(-A_+,-A_-,-2j;0;-(\alpha_++\zeta)(\alpha_-+\zeta^{\ast})e^{\alpha_0})}{_3F_0(-A_+,-A_-,-2j;0;-|\zeta|^2)} \ ,
\eea
where $A_+$ and $A_-$ are as defined in section IV. In the above
derivation we have made use of the identity
\beq
{e^{\alpha
Z}X_+e^{-\alpha Z}=e^{\alpha}X_+}.
\eeq
We will use this identity in
simplifying the symbol of the product of two general operators
labeled by $\hat{\alpha}$ and $\hat{\beta}$.

We now compute the following to find the star product in terms of
deformations of the point-wise product:
\beq{\phi(\hat{\alpha}\hat{\beta})=N^{-2}\langle\zeta|e^{\alpha_-X_-}e^{\alpha_0
Z}e^{\alpha_+X_+}e^{\beta_-X_-}e^{\beta_0
Z}e^{\beta_+X_+}|\zeta\rangle} \ .\eeq
We give the final result of this
matrix element without going through the steps;
\bea \nonumber
\phi(\hat{\alpha}\hat{\beta}) & = &
N^{-2}\phi(\hat{\alpha})\phi(\hat{\beta})+N^{-2}e^{-j(\alpha_0+\beta_0)} \left[\phi(\hat{\alpha})e^{j\alpha_0}
+\chi(\hat{\alpha})e^{j\alpha_0} \right. \\ \label{Hstar} & + & \left.
e^{j(\alpha_0+\beta_0)}\{\phi(\hat{\alpha})\phi(\hat{\beta})+\chi(\hat{\alpha})\chi(\hat{\beta})
+\phi(\hat{\alpha})\chi(\hat{\beta})+\chi(\hat{\alpha})\phi(\hat{\beta})\} +  L \right] \ .
\eea
In this expression,
\bea
\chi(\hat{\alpha})=e^{-j\alpha_0}\sum_{i=1}^{2j}\frac{(\alpha_-+\zeta^{\ast})^i(-\zeta)^ie^{i\alpha_0}}{(i!)^2}\prod_{l=0}^{i-1}K_{j,-j+l}^2\sum_{k=0}^{i-1}\frac{(\alpha_++\zeta)^k}{(-\zeta)^k}
\left({\begin{array}{*{10}c} i \\ k \\ \end{array}}\right), \\
\chi(\hat{\beta})=e^{-j\beta_0}\sum_{i=1}^{2j}\frac{(\beta_++\zeta)^i(-\zeta^{\ast})^ie^{i\beta_0}}{(i!)^2}\prod_{l=0}^{i-1}K_{j,-j+l}^2\sum_{k=0}^{i-1}\frac{(\beta_-+\zeta^{\ast})^k}{(-\zeta^{\ast})^k}\left({\begin{array}{*{10}c} i \\ k \\ \end{array}}\right),
\eea
and
\bea
L & = &
\sum_{i=1}^{2j}\frac{(\beta_++\zeta)^i(\alpha_-+\zeta^{\ast})^ie^{i(\alpha_0+\beta_0)}}{(i!)^2}
\prod_{l=0}^{i-1}K_{j,-j+l}^2 \bigg[
{_3}F_0(-A_+,-A_-,i-2j;0;-\alpha_+(\alpha_-+\zeta^{\ast})e^{\alpha_0})
\\ \nonumber & + &
\sum_{m=1}^{i-1}\frac{\beta_-^m}{(\alpha_-+\zeta^{\ast})^me^{m\alpha_0}}\left(^i_m\right)
{_3}F_0(-A_+,-A_-,i-2j-m;0;-\alpha_+(\alpha_-+\zeta^{\ast})e^{\alpha_0}) \bigg]+1.
\eea
The computations involve some non-trivial simplifications to
bring it to Eq. (\ref{Hstar}).

We see that the first term in Eq. (\ref{Hstar}) is the point-wise
product of the the two symbols and the term in the bracket gives the
deformations.

As the star product was computed using the HACS we can be sure that
they are well behaved at the conical singularity seen in the
continuum. The reason is same as mentioned in section III.

\section{conclusions}

In the present paper we have analyzed algebraic structures that are
more general than the fuzzy sphere. We have shown that the
interesting feature of the topology change studied in \cite{Arlind}
also occurs in the present case. It must be mentioned here that the
HA arises naturally in the study of integrable dynamics of two
dimensional curved surfaces. Study of nonlinear deformations of
algebra and their representation theory assumes importance in the
context of noncommutative geometries. There have been attempts to
write down the Dirac operator for the $SU_q(2)$ algebra
\cite{Dabrowski,Harikumar}. Such attempts can be extended to this
nonlinear algebra too. Quantum field theories on fuzzy spheres
coming from $SU(2)$ representations can be further extended to  our
framework also. Topology change will play an important role in such
studies which will be explored in detail. Fuzzy spheres are
considered in the Kaluza-Klein framework and similar studies for
Higgs algebra will bring out new features. These novel surfaces can
also be generated dynamically along the lines of fuzzy spheres and
they will naturally arise when one introduces higher dimensional
operators in the effective action of QFT's.

\acknowledgments

We thank Prof. A. P. Balachandran for many useful discussions and
constant encouragement. PP thanks the Director IMSc for providing
visitorship at the institute. The work was supported in part by DOE
under the grant number DE-FG02-85ER40231. TRG would like to thank Prof.
Hermann Nicolai for support at AEI, Golm.

\end{document}